\documentclass[aps,pre,twocolumn,galleyproof,superscriptaddress]{revtex4-1}
\usepackage{amsmath}
\usepackage{graphicx,color,dsfont}
\usepackage{etoolbox}

\begin{document}

\title{Inferring human mobility using communication patterns}
\date{\today}

\author{Vasyl Palchykov}
\affiliation{Department of Biomedical Engineering and Computational Science (BECS), Aalto University School of Science, P.O. Box 12200, FI-00076, Finland}
\affiliation{Institute for Condensed Matter Physics, National Academy of Sciences of Ukraine, UA 79011 Lviv, Ukraine}
\affiliation{Lorentz Institute, Leiden University, 2300 RA Leiden, The Netherlands}
\author{Marija Mitrovi\'c}\affiliation{Department of Biomedical Engineering and Computational Science (BECS), Aalto University School of Science, P.O. Box 12200, FI-00076, Finland}
\affiliation{Scientific Computing Laboratory, Institute of Physics Belgrade, University of Belgrade, Pregrevica 118, 11080 Belgrade, Serbia}
\author{Hang-Hyun Jo}
\affiliation{Department of Biomedical Engineering and Computational Science (BECS), Aalto University School of Science, P.O. Box 12200, FI-00076, Finland}
\affiliation{BK21plus Physics Division and Department of Physics, Pohang University of Science and Technology, Pohang 790-784, Republic of Korea}
\author{Jari Saram\"aki}\affiliation{Department of Biomedical Engineering and Computational Science (BECS), Aalto University School of Science, P.O. Box 12200, FI-00076, Finland}
\author{Raj Kumar Pan}\affiliation{Department of Biomedical Engineering and Computational Science (BECS), Aalto University School of Science, P.O. Box 12200, FI-00076, Finland}

\begin{abstract}
  Understanding the patterns of mobility of individuals is crucial for a number of reasons, from city planning to disaster management. There are two common ways of quantifying the amount of travel between locations: by direct observations that often involve privacy issues, e.g., tracking mobile phone locations, or by estimations from models. Typically, such models build on accurate knowledge of the population size at each location. However, when this information is not readily available, their applicability is rather limited. As mobile phones are ubiquitous, our aim is to investigate if mobility patterns can be inferred from aggregated mobile phone call data alone. Using data released by Orange for Ivory Coast, we show that human mobility is well predicted by a simple model based on the frequency of mobile phone calls between two locations and their geographical distance. We argue that the strength of the model comes from directly incorporating the social dimension of mobility. Furthermore, as only aggregated call data is required, the model helps to avoid potential privacy problems.
\end{abstract}
\maketitle


\section*{Introduction}
People travel and move for a variety of reasons, including social, economic, and political factors. While individuals may follow simple, recurrent patterns of movement, e.g., daily commuting, a more complex picture emerges when all trajectories of a population are assembled together~\cite{brockmann_scaling_2006}.
Understanding the principles governing individual and collective movement is important for a number of reasons: for planning urban design~\cite{hall_cities_2002}, for forecasting and avoiding traffic congestion~\cite{helbing_traffic_2001}, for mitigating infectious disease~\cite{Balcan2009,Wesolowski2012,Dalziel2013}, and for contingency planning in extreme situations caused by disasters~\cite{helbing_simulating_2000,Lu2012}. However, accurately determining the movement patterns in a population is cumbersome and costly, and involves privacy issues. 

There are two ways of inferring the mobility patterns in a population: by direct measurement or by models that predict population movement based on other observed data. Regarding the former, tracking the movement of individuals using location data from mobile phones~\cite{Gonzalez2008,Song2010,Jo2012} has emerged as a powerful alternative to traditional methods such as traffic surveys~\cite{Treiterer1975}. In this case, the data set comes from the billing systems of mobile phone operators, where the closest tower of each phone is recorded when a mobile phone is used. The resolution problems caused by this are compensated by the large quantity and high quality of data~\cite{Calabrese2011,Tizzoni2013}. However, there are drawbacks to this approach: tracking the locations of individuals may be seen as a threat to privacy even when the data is properly anonymised~\cite{Butler2007}.

The alternative approach to direct measurement is to use models that predict the average population behaviour from (publicly) available information, such as census and population data. Perhaps the most famous example is the gravity model~\cite{Carey1867,Carrothers1956,anderson_gravity_2011} that has been used to predict the intensity of a number of  human interactions, including population movement~\cite{barthelemy_spatial,jung_epl,thiemann_plos_one} and mobile phone calls between cities~\cite{krings_urban_2009}. In the gravity model, the intensity of interactions between two locations (e.g., cities) is determined by their populations and distance (with proper scaling exponents). Recently, it has been shown that a parameter-free model, the radiation model~\cite{Simini2012}, is able to predict mobility patterns with improved accuracy; this model requires geospatial information on population size as an input. 

The applicability of the above-mentioned models is constrained by the availability of accurate population information. This may become a problem e.g. for developing countries, where census data may be incomplete. However, mobile phones are ubiquitous almost everywhere, and one might expect that mobile phone calls reflect the social dimension of mobility -- the amount of social ties between geospatial locations can be expected to influence travel patterns. Therefore, the aim of this paper is to predict mobility patterns from mobile phone call data alone, and examine models that would be applicable in a setting where accurate, up-to-date population information is not available. Furthermore, we focus on models that only require aggregated call data, without needing to track individual users. This has the obvious benefit of mitigating privacy-related issues; additionally, the volume of required input data is smaller and the aggregation can be easily done by the mobile operator that owns the source data. 

Our modelling and analysis is purely based on the Ivory Coast mobile telephone data set~\cite{d4d}, originally released by Orange for the Data for Development Challenge. This data set includes information on mobile phone calls aggregated at the tower level during 140 days, used as inputs for the models, and data on the trajectories of randomly chosen individuals, used for developing the models and testing their accuracy. There is no accurate, up-to-date geospatial population information for Ivory Coast; the last census was conducted in 1998, and there is no data available on mobility or migration within the country. In contrast, the telephone system in Ivory Coast is well-developed by African standards with mobile phone penetration above 83\%~\cite{budde}.

This paper is constructed as follows: first, we examine gravity laws for average mobility and call frequency between locations. We then proceed to show that mobility between two locations can be directly estimated from the number of calls between the locations and their distance. This holds at two levels of coarse-graining: between tower locations in a major city and between cities. Finally, we study the accuracy of predictions for individual pairs of locations, beyond averages, and show that the number of calls between locations appears to be a good predictor of the frequency of travel between them. For reference, we also study variants of existing mobility models (the gravity and radiation models) where location-specific call frequencies are used as inputs instead of population data; despite applying these models beyond their intended range, they provide fairly good predictions on average. 

\section*{Results}
\subsection*{Data set and coarse-graining}

The data set comes in two parts: (i) the number of calls between 1231 Orange towers in Ivory Coast for 5 months, and (ii) ten data sets on two-week individual trajectories of 50,000 randomly chosen users. From the trajectories, we aggregated the mobility $m_{ij}$ between locations $i$ and $j$ by counting direct movements along the trajectories (see Methods for further details). 

As it is reasonable to assume that communication and mobility patterns are in general different for short and long distances, we aggregated the data at two levels: (i) tower level for intra-city behaviour and (ii) city level for inter-city behaviour. The intra-city analysis consist of 5.1 million movements and 109 million calls between all 298 towers located inside Abidjan, the largest city of Ivory Coast, during 140 days. This comprises  $31\%$ of all calls and $50\%$ of all movements in the country. In this analysis the geographical unit  -- referred to as ``location'' in the following -- is the area covered by a single tower. To analyse inter-city behaviour, we aggregated towers that lie within a city boundary and consider calls and mobility between cities. The resulting data contains 143 cities with 63 million calls and $374$ thousand movements between them during 140 days. At both levels of analysis, we determine the number of calls, movements, and the geographical distance between every pair of locations (towers, cities). See Methods for further details.


\subsection*{Gravity laws: dependence of mobility and communication intensity on distance}

We begin by investigating whether the mobility and communication intensities between two locations follow the gravity law  on average. In its general form, the gravity law states that
\begin{equation}
x_{ij} \propto \frac{N_i N_j}{d_{ij}^\alpha},
\end{equation}
where $x_{ij}$ is the intensity of interaction, e.g., calls, mobility, trade, between locations $i$ and $j$ associated with populations of sizes $N_i$ and $N_j$, separated by a distance $d_{ij}$~\cite{Carey1867,Carrothers1956,anderson_gravity_2011}. The exponent $\alpha$ governs the distance dependence. Note that in the most general form of the gravity law, $N_i$ and $N_j$ are also associated with an exponent; here for simplicity we assume a linear dependence. For our data, we study the intensities of mobility $m_{ij}$ and communication $c_{ij}$ between locations $i$ and $j$. These are defined as the average number of weekly movements and calls between them, respectively.  As a proxy of the population $N_i$, we take the total number of weekly calls $s_i$ made and received at location $i$. 

\begin{figure}[!t]
\includegraphics[width=8.5cm]{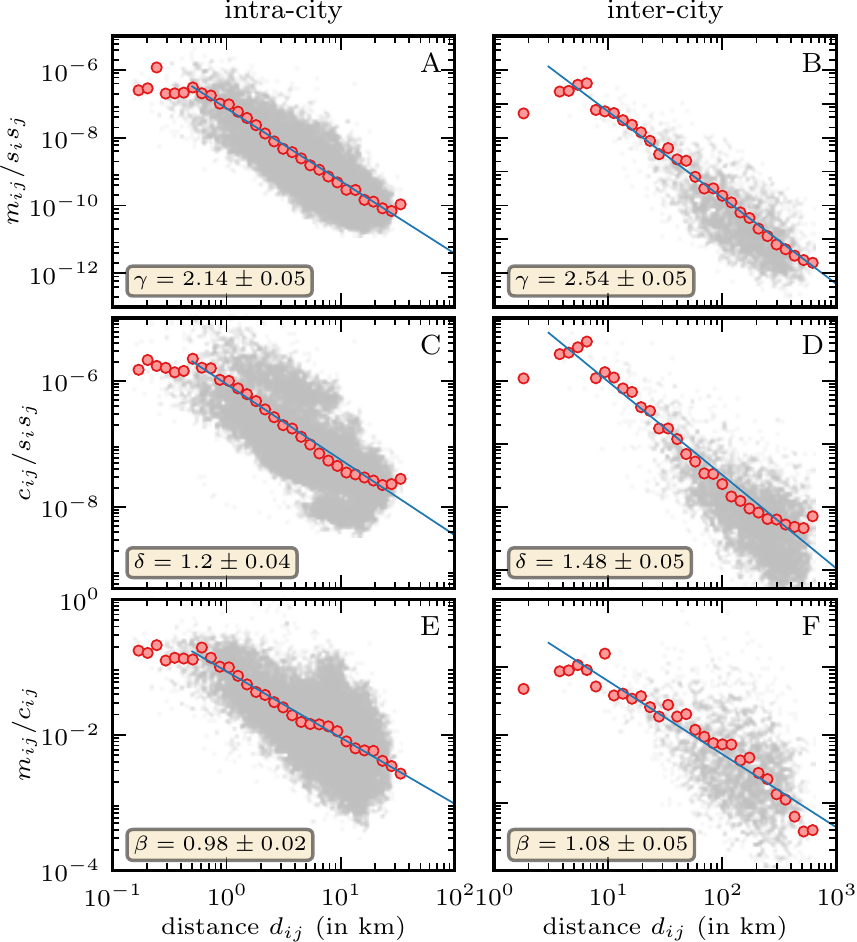}
\caption{Dependence of the intensities of interaction on distance. The number of (A,B) movements per strength product $m_{ij}/s_is_j$, (C,D) calls per strength product $c_{ij}/s_is_j$, and (E,F) movements per call $m_{ij}/c_{ij}$ decrease with distance between $i$ and $j$ for both intra-city and inter-city analyses. Each grey dot indicates a pair of locations, and circles correspond to the average log-binned behaviour. Solid lines show the fitted power-law decaying behaviour.}
\label{fig:fig1}
\end{figure}

The variation of the scaled mobility intensity, $m_{ij}/s_is_j$, with respect to the distance $d_{ij}$ is shown  in Fig.~\ref{fig:fig1} for the tower and city levels of coarse-graining (panels A and B, respectively). In both cases, the gravity law holds on average and
\begin{equation}
  \left\langle{\frac{m_{ij}}{s_is_j}}\right\rangle \propto d_{ij}^{-\gamma}, \label{eq:mobgravity}
  \end{equation}
 where $\gamma\approx 2.14$ for the intra-city level and $\gamma \approx 2.54$ for the inter-city level. Panels C and D display a similar plot for the scaled communication intensity that is also seen on average to follow the gravity law:
 \begin{equation}
  \left\langle{\frac{c_{ij}}{s_is_j}}\right\rangle \propto d_{ij}^{-\delta}, \label{eq:callgravity}
  \end{equation}
  where the distance exponents are $\delta \approx 1.20$ for the intra-city level and $\delta \approx 1.48$ for the inter-city level. It is worth noting that both exponents $\gamma$ and $\delta$ are smaller for the intra-city level, indicating differences in communication and travel patterns within and between cities: within a city, the spatial distance appears to play a less important role than it does between cities.
  
The two gravity laws discussed above suggest that the following relationship might also hold:
  \begin{equation}
\left\langle \frac{m_{ij}}{c_{ij}}\right\rangle \propto d_{ij}^{-\beta},\label{eq:mobcall}
  \end{equation}
  where $\beta=\gamma - \delta$. This is indeed the case, as seen in Fig.~\ref{fig:fig1} (E,F) where $\left\langle m_{ij}/c_{ij}\right\rangle$ follows a power-law dependence on $d_{ij}$. For both intra- and inter-city levels, we find the exponent $\beta \approx \gamma - \delta$ (see Table~\ref{tab:tab1}). These results suggest that there are two possible ways of inferring the intensity of mobility between locations $i$ and $j$ from call data: using the distance and either (i) the total call numbers at both locations $s_i$ and $s_j$ (Eq.~\ref{eq:mobgravity}), or (ii) the total number of calls between the locations $c_{ij}$ (Eq.~\ref{eq:mobcall}). The prediction accuracy of these two models will be assessed in in the section "Prediction accuracy" below.  

\begin{table}[!t]
\centering
\begin{tabular}{l|c|c|c}
\hline\hline
Level & $\gamma$ & $\delta$ & $\beta$\\ \hline
intra-city (tower level)          & $2.14\pm 0.05$ & $1.20\pm0.04$ & $0.98\pm0.02$\\
inter-city (city level)              & $2.54\pm 0.05$ & $1.48\pm0.05$ & $1.08\pm0.05$\\ 
\hline\hline
\end{tabular}
\caption{The estimated values of exponents $\gamma$ (Eq.~\ref{eq:mobgravity}), $\delta$ (Eq.~\ref{eq:callgravity}), and $\beta$ (Eq.~\ref{eq:mobcall}) for the tower and city levels of coarse-graining. The values and their standard errors have been obtained by least square fitting to logarithmically binned data.}
\label{tab:tab1}
\end{table}

It is worth noting that both for intra- and inter-city levels, the exponent $\beta \approx 1$. This does not directly result from Eqs.~(\ref{eq:mobgravity}) and (\ref{eq:callgravity}). One possible argument for the observed value of $\beta$ is as follows: the cost of a single trip, measured in e.g. time or money, between two towers/cities $i$ and $j$ can be assumed to depend linearly on their distance, $d_{ij}$. This means that the total cost of all movements between $i$ and $j$ is proportional to $m_{ij}d_{ij}$. However, the cost of communication is independent of distance. If one further assumes that the total cost of movement is balanced by the total benefit brought by social ties, linearly reflected in $c_{ij}$, we have $m_{ij}d_{ij}\sim c_{ij}$ and thus the value of exponent $\beta=1$. In this interpretation, the communication exponent $\delta$ is directly related to a decrease in the number of social ties as function of distance, whereas $\gamma$ captures a combination of cost associated with travel and the decrease in the number of social ties.

\subsection*{Models for estimating mobility based on call data}

The results of the previous section indicate that on average, the mobility intensity $m_{ij}$ between two locations $i$ and $j$ can be estimated using the \emph{gravity model}
\begin{equation}
m_{ij}^\mathds{G}=k^\mathds{G}\frac{s_is_j}{d_{ij}^{\gamma}},\label{eq:model_gravity}
\end{equation}
where $k^\mathds{G}$ is a normalization constant obtained by equating the total numbers of expected and observed movements, i.e., $\sum_{ij} m_{ij}=\sum_{ij} m_{ij}^\mathds{G}$. This model takes  the communication intensities $s_i$ and $s_j$ at both locations as inputs in addition to the distance $d_{ij}$. As an alternative we propose the \emph{communication model}
\begin{equation}
 m_{ij}^\mathds{C}=k^\mathds{C}\frac{c_{ij}}{d_{ij}^{\beta}}, \label{eq:model_communication}
 \end{equation}
 based on the communication intensity $c_{ij}$ between the locations. The normalization constant $k^\mathds{C}$ is obtained as before. The values of the exponents $\gamma$ and $\beta$ are taken from Table~\ref{tab:tab1}.
 
For comparison, we also study a modified version of the \emph{radiation model}~\cite{Simini2012}, originally designed to predict mobility between locations $i$ and $j$ with the help of data on population density in the surrounding area. Again, we modify the model such that only call and distance data is required as input. To this end, we assume that the number of calls in a given location is an unbiased estimate of population density, similarly to the gravity model. Note that this assumption may not necessarily hold, since mobile phone penetration may correlate with socioeconomic factors.  Further, we assume that the number of trips that begin (end) at location $i$ ($j$) is proportional to $s_i$ ($s_j$). Then, the radiation model formula can be rewritten as
 \begin{equation}
\begin{aligned}
  \label{eq:eq5}
  m_{ij}^\mathds{R} &= k^\mathds{R} \left[\frac{s_i^2 s_j}{(s_i+s_{ij})(s_i+s_j+s_{ij})}\right.\\
  &+ \left. \frac{s_i s_j^2}{(s_j+s_{ji})(s_j+s_i+s_{ji})} \right].
\end{aligned}
\end{equation}
Here $s_{ij}$ denotes the total number of calls made within a circle of radius $d_{ij}$ centred at $i$, excluding locations $i$ and $j$, and $k^\mathds{R}$ is a normalization constant. 

\subsection*{Prediction accuracy}\label{sec:accuracy}
\begin{figure}[!t]
\includegraphics[width=9.0cm]{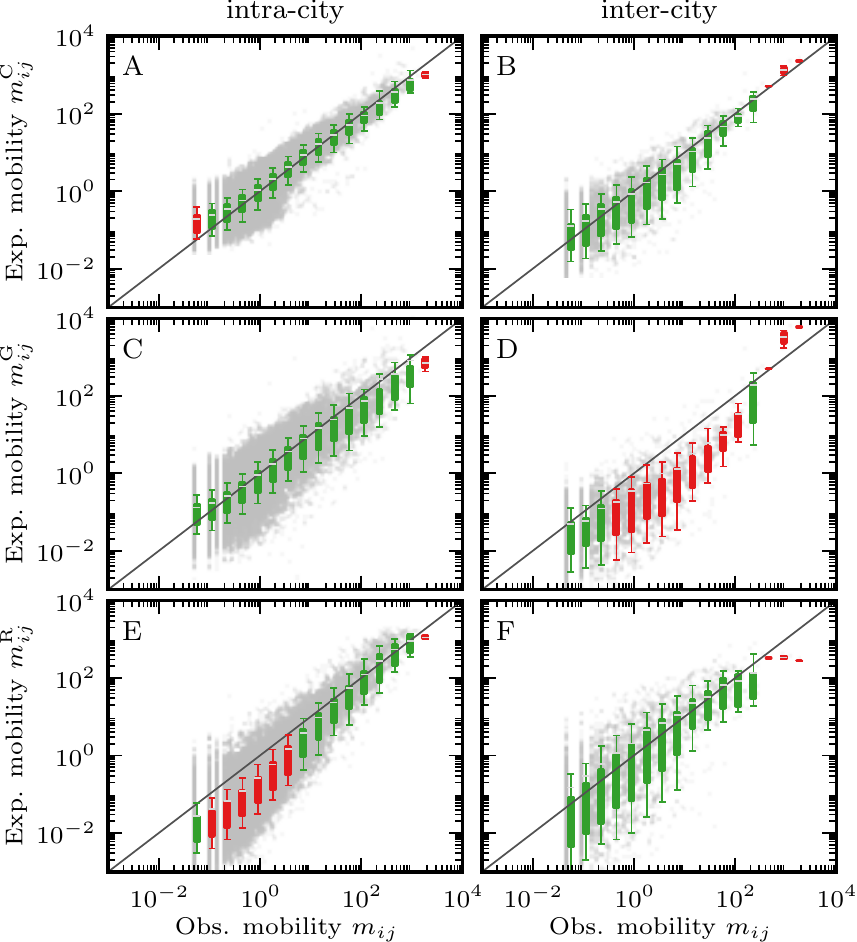}
\caption{Comparison between observed and predicted human mobility. The expected mobility intensities (A,B) $m_{ij}^\mathds{C}$ for the communication model, (C,D) $m_{ij}^\mathds{G}$ for the gravity model, and (E,F) $m_{ij}^\mathds{R}$ for the radiation model are plotted against the mobility intensities observed in data $m_{ij}$. The left panels (A,C,E) correspond to the intra-city analysis and right panels (B,D,F) correspond to inter-city analysis. The boxes provide the region between 25th and 75th percentiles, and the whiskers correspond to 9th and 91st percentiles of logarithmically binned data. A box is colored green if for a given bin the line $y=x$ lies between the 9th and the 91st percentiles of the expected distribution; otherwise it is colored red.}
\label{fig:fig2}
\end{figure}

To assess the actual predictive power of the models beyond averages, we compare the actual mobility intensity $m_{ij}$, obtained from the trajectory data set, with the estimates given by the models for each specific pair of locations $i$ and $j$. This comparison for the communication model, the gravity model, and the radiation model is shown in Fig.~\ref{fig:fig2}. The gray dots correspond to predicted versus actual mobility for each pair of locations, and the boxes (whiskers) correspond to the region between 25th and 75th (9th and 91st) percentiles.

\begin{table}[t]
\centering
\begin{tabular}{l|c|c|c|c|c}
\hline\hline
Level                & $r_s^\mathds{C}$ & $r_s^\mathds{G}$ & $r_s^\mathds{R}$ &  $p$ ($r_s^\mathds{C}>r_s^\mathds{G}$) & $p$ ($r_s^\mathds{C}>r_s^\mathds{R}$)\\ \hline
intra-city (tower level) & 0.87 & 0.81 & 0.82 & $<10^{-4}$ & $<10^{-4}$\\ 
inter-city (city level)  & 0.74 & 0.67 & 0.67 & $<10^{-4}$ & $<10^{-4}$\\
\hline\hline
\end{tabular}
\caption{Spearman correlation coefficient between the observed and predicted mobility values for the three models. For both intra-city and inter-city analyses the communication model shows larger correlation values than gravity and radiation models. The significance of the difference in the correlation is indicated by the $p$-values.}
\label{tab:tab_2}
\end{table}

It is clear from the figure that all models give on average reasonable predictions. However, the gravity and radiation models display higher levels of variance between the predicted and actual mobility intensities. In particular, the prediction accuracy of the gravity model is relatively poor for the inter-city mobility, and the radiation model performs the worst for the intra-city mobility. The latter is not surprising, as the radiation model was originally not designed for predicting short-range travel patterns within cities. Further, the original radiation model requires accurate geospatial population information, and simply equating population size within an area with the number of calls can be expected to give rise to errors.

The level of observed variance implies that in addition to comparing averages, it is important to compare the expected and observed mobility between individual pairs of locations. As the first step, we determine the Spearman correlation coefficients $r^\mathds{C,G,R}$ between $m_{ij}$ and $m_{ij}^\mathds{C,G,R}$. Table~\ref{tab:tab_2} shows that the correlation is higher for the communication model than for the gravity and radiation models for both levels of coarse-graining (intra-city, inter-city). In general, in terms of the Spearman coefficient, predictions of all models are more accurate for intra-city mobility than for inter-city mobility.

Finally, we consider the differences between the observed and  predicted mobilities by measuring their relative deviations. For all the three models, we define the relative deviations $\delta_{ij}^\mathds{C,G,R}$ between the observed $m_{ij}$ and predicted $m_{ij}^\mathds{C,G,R}$ as
\begin{equation}\label{eq:reldev}
\delta_{ij}^\mathds{C,G,R}=\frac{m_{ij}^\mathds{C,G,R}-m_{ij}}{m_{ij}^\mathds{C,G,R}+m_{ij}},
\end{equation}
where $\delta_{ij}$ takes values between $-1$ and $1$. A deviation of $\delta_{ij}=0$ implies exact prediction by the model for the pair of locations $i$ and $j$, whereas negative (positive) values indicate under- (over-) estimations. We  only determine $\delta_{ij}$ for those pairs of of $i$ and $j$ for which $m_{ij}\neq0$.

\begin{figure}[!t]
\includegraphics[width=9.0cm]{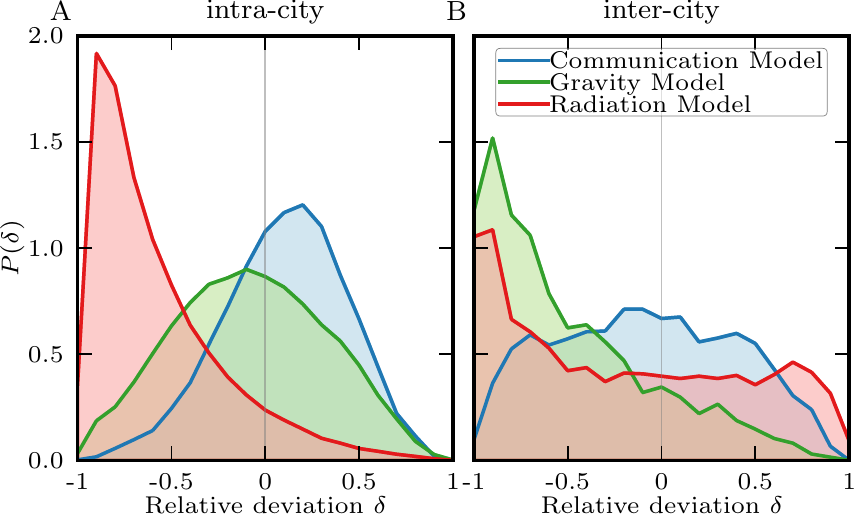}
\caption{Relative deviation between the observed and predicted mobility values for the three models. Distribution $P(\delta_{ij}^\mathds{C,G,R})$ of the relative deviations $\delta_{ij}^\mathds{C,G,R}$ (Eq.~\ref{eq:reldev}) for (A) intra-city and (B) inter-city mobility.}
\label{fig:fig3}
\end{figure}

The probability distributions $P(\delta^\mathds{C,G,R})$ shown in Fig.~\ref{fig:fig3} confirm the above finding that out of the studied three models for inferring mobility from call data, the communication model has the highest accuracy of prediction. The distribution $P(\delta^\mathds{C})$ is well centred around zero, whereas especially for inter-city mobility the distributions $P(\delta^\mathds{G})$ and $P(\delta^\mathds{R})$ show a bias towards under-estimation. In more detail, for intra-city mobility, the fractions of location pairs with deviations $\delta \in [-0.25,0.25]$ are 13\% for the radiation model, 42\% for the gravity model, and 51\% for the communication model. 
 For  inter-city mobility, the corresponding fractions are 20\%, 17\% and 33\%. Note that for the gravity model, in spite of the fact that the average $\left<m_{ij}/(s_is_j)\right>$ follows a $d_{ij}^{-\gamma}$-dependence (Fig.~\ref{fig:fig1} A,B), there is still a significant amount of under-estimation. This indicates that there is a broad distribution of the values of $\left<m_{ij}/(s_is_j)\right>$ for a given distance, and the average value is not always a good estimator.  

\section*{Discussion and conclusion}
The goal of this paper has been to investigate simple models that predict the intensities of mobility between two locations on the basis of mobile phone call data and their geospatial distance. The motivation behind this is to provide ways of predicting mobility in situations where accurate information of population size at each location is not available; furthermore, the focus is on aggregated call data, mitigating the need to track movement patterns of individual phone users. Our study is based on call and mobility data released by Orange for Ivory Coast; note that it would be important to verify the findings with data from other countries.

We have tested three models that only take aggregated call data and geospatial information as inputs: the well-known gravity model, the communication model based on the number of calls between two locations, and a modified version of the radiation model. While all models on average capture the real mobility patterns derived from call data with location information, a more detailed analysis of the prediction accuracy at the level of individual locations reveals that the communication model is the most accurate out of the three tested models in this setting. 

Note that the gravity and radiation models were originally designed to use geospatial population information as input parameters. Since our aim has been to study mobility models in a setting where such information is not available, we have simply taken the number of calls at a given location as a proxy of the population size. Therefore we do not claim that the communication model would outperform other models in a situation where they could be applied as their designers intended. Also note that our modeling target -- the mobility pattern -- is also derived from mobile phone records, and geospatial biases in mobile phone usage might influence the results. Hence, it would be useful to verify the accuracy of the communication model for a case where there are alternative sources of mobility information.

The likely reason why the communication model works well is that it directly incorporates geospatial information on social ties and human relationships. It has been observed earlier that individuals tend to travel to locations where they have social bonds~\cite{Lu2012}; furthermore, once under way, it is reasonable to assume that people make calls back home. Because of this, the aggregated intensity of communication between two locations should contain information on the mobility patterns as well. Then, in the first approximation one might assume that the frequency of movement between two locations is directly proportional to the intensity of communication. Further, the simplest way to incorporate the fact that larger distances imply larger travel costs (in terms of time or money) is to assume that mobility is inversely proportional to distance. These two components directly yield the communication model: $m_{ij}\propto c_{ij}/d_{ij}$.

It is worth noting that in general, in gravity laws of human interaction, the distance dependence is associated with some exponent $\alpha$. This is also seen in our analysis of the gravity laws for mobility and communication intensity, where the exponents were seen to depend on the level of coarse-graining, i.e., intra-city or inter-city. However, for both levels, the inverse distance dependence of the communication model is approximately linear, i.e., the exponent equals one. This suggests universality and calls for analysis of similar data sets from different countries.

\section*{Methods}

\subsection*{Communication and mobility data}
The data set \cite{d4d} consists of 2.5 million call detail records of customers for a single provider (Orange) in Ivory Coast between December 1st, 2011 and April 28th, 2012. The communication data used in this paper contains the number of calls as well as their aggregated duration between all pairs of 1231 towers, i.e., mobile base stations. The geographical locations of the towers were also provided. The temporal resolution of the data set is one hour.

The mobility sample consists of ten data sets of trajectories of individual users, each for 50,000 randomly chosen users. Each trajectory corresponds to the subscribers' call locations during a two-week period. The locations were recorded every time a call was made and correspond to the position of the tower that transmitted the call. The data sets represent consecutive two-week periods, beginning in December 5, 2011.

\subsection*{Determining city boundaries}
As the locations of the cell-towers were provided, we used reverse geocoding~\cite{google} to determine the city in which the tower is located. The mean longitude and latitude of all towers within a city defines the centre of the city. This location was used to calculate the inter-city distances. Out of the 1231 mobile phone towers, 686 are located within city boundaries (with 298 of them in the largest city, Abidjan). The total number of cities with at least a single tower is 143.

\subsection*{Determining direct movements}
Given the individual trajectories of users, a variety of methods have been developed to extract different aspects of human mobility~\cite{Calabrese2011}. Here, we consider \textit{direct movements} that correspond to any consecutive changes in the location of a user. Formally, direct movements are defined as follows: if the user made a call from location $i$ at some time $t$ and $j$ is the location of the next call at $t'>t$, there is a direct movement from $i$ to $j$ if $j \neq i$.  By aggregating this information for all users we determine, the total number of direct movements between all pairs of locations. The locations can correspond either to towers (intra-city analysis) or to cities (inter-city analysis). Note that for inter-city analysis, only towers located within city boundaries are considered. Thus, all  calls and direct movements to locations between cities are ignored. 

\subsection*{Data filtering}
Users may be located in areas covered by several towers. In this case, the calls made by users at the same location can be handled by different neighbouring towers. This phenomena of switching of mobile phone calls between towers is called \textit{handover} and it may give rise to artefacts in mobility and communication. For instance, let us consider an \textit{immobile} user located in the boundary area covered by two towers $i$ and $j$. If one of the calls of this user was served by tower $i$ and the subsequent call by tower $j$, the data will indicate movement of the user from tower $i$ to tower $j$. Similarly, the number of calls between neighbouring towers might also get biased. To get rid of this artefact, we excluded all pairs of neighbouring towers from our analysis. As the towers are heterogeneously distributed (higher concentration in densely populated areas and lower concentration in rural zones), neighbouring towers were identified by a distance-independent approach. To do this, we first computed the Voronoi diagram around each tower. The towers having a common edge in their Voronoi cells are defined as the neighbouring towers. We also excluded the communication and mobility between the towers that are located within 1 meter from each other (e.g. two base stations serving a busy area). Further, only pairs of locations with more than one call per day (on average) were considered. 

\section*{Acknowledgements}
We thank the operator France Telecom-Orange and the ``Data for Development'' committee for sharing the mobile phone dataset and organizing the D4D challenge. We acknowledge the support by the Academy of Finland, project no. 260427 (JS, RKP) and Aalto University postdoctoral program (HJ). VP was supported by TEKES (FiDiPro). MM was supported in part by the Ministry of Education, Science, and Technological Development of the Republic of Serbia under project no. ON171017. We also acknowledge the computational resources provided by Aalto Science-IT project.

\def\url#1{}

\clearpage

\end{document}